\documentclass[acmsmall, nonacm]{acmart} 

\usepackage{subcaption}

\AtBeginDocument{%
  \providecommand\BibTeX{{%
    \normalfont B\kern-0.5em{\scshape i\kern-0.25em b}\kern-0.8em\TeX}}}

\acmConference[ETRA '22]{ETRA '22: ACM Symposium on Eye Tracking Research and Applications}{June 08--11, 2022}{Seattle, WA}
\acmBooktitle{ETRA '22: ACM Symposium on Eye Tracking Research and Applications,
  June 08--11, 2022, Seattle, WA}
\acmPrice{15.00}
\acmISBN{978-1-4503-XXXX-X/18/06}

\graphicspath{{figures/}{pictures/}{images/}{./}}

\begin{document}

\title[A Spiral into the Mind]{A Spiral into the Mind: Gaze Spiral Visualization for Mobile Eye Tracking}


\author{Maurice Koch}
\authornotemark[1]
\email{maurice.koch@visus.uni-stuttgart.de}
\affiliation{%
  \institution{University of Stuttgart}
  \city{Stuttgart}
  \country{Germany}
}

\author{Daniel Weiskopf}
\affiliation{%
  \institution{University of Stuttgart}
  \city{Stuttgart}
  \country{Germany}
}
\email{daniel.weiskopf@visus.uni-stuttgart.de}

\author{Kuno Kurzhals}
\email{kuno.kurzhals@visus.uni-stuttgart.de}
\affiliation{%
  \institution{University of Stuttgart}
  \city{Stuttgart}
  \country{Germany}
}

\renewcommand{\shortauthors}{Koch et al.}

\begin{abstract}
Comparing mobile eye tracking data from multiple participants without information about areas of interest (AOIs) is challenging because of individual timing and coordinate systems. 
We present a technique, the gaze spiral, that visualizes individual recordings based on image content of the stimulus. 
The spiral layout of the slitscan visualization is used to create a compact representation of scanpaths. 
The visualization provides an overview of multiple recordings even for long time spans and helps identify and annotate recurring patterns within recordings. 
The gaze spirals can also serve as glyphs that can be projected to 2D space based on established scanpath metrics in order to interpret the metrics and identify groups of similar viewing behavior. 
We present examples based on two egocentric datasets to demonstrate the effectiveness of our approach for annotation and comparison tasks. 
Our examples show that the technique has the potential to let users compare even long-term recordings of pervasive scenarios without manual annotation.
\end{abstract}



%\setcopyright{acmlicensed}
%\acmJournal{PACMCGIT}
%\acmYear{2022} \acmVolume{5} \acmNumber{2} 
%\acmArticle{etra-fp1000} \acmMonth{6} 
%\acmPrice{15.00}\acmDOI{10.1145/3530795}


\begin{teaserfigure}
\centering
\includegraphics[width=0.8\linewidth]{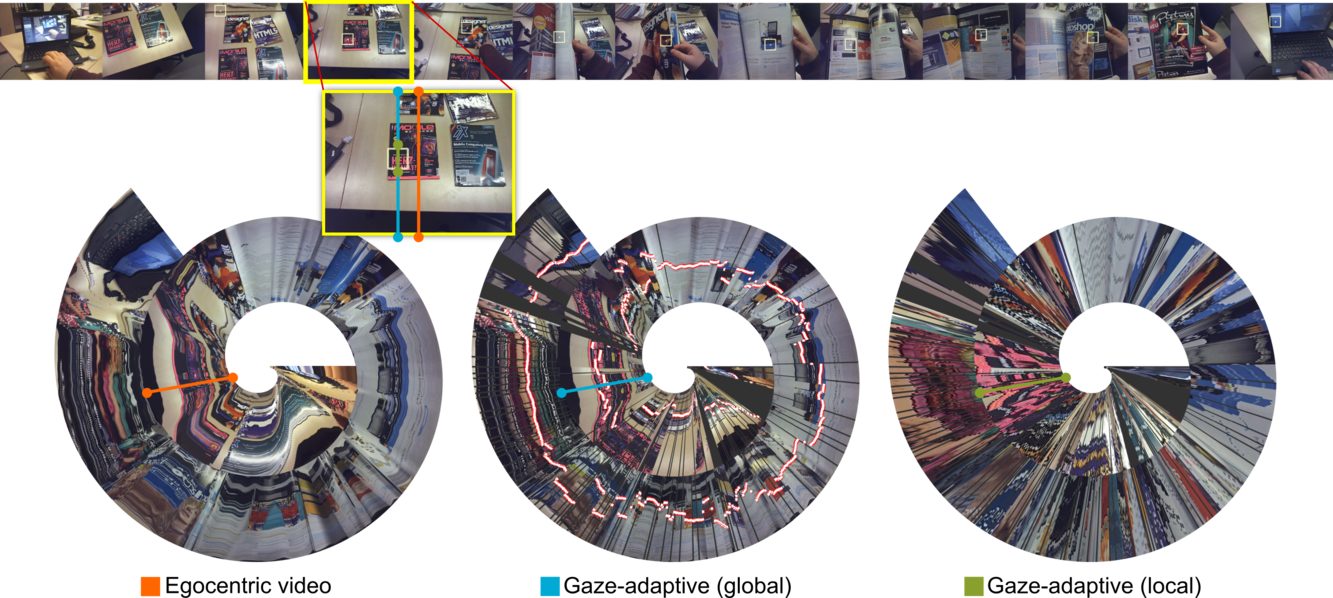}
\caption{Video and gaze data of a person looking at different magazines (0:40 min). For each frame, a vertical scanline is extracted to create the visualization. The location and size of the scanline is either constant (orange) or influenced by the current point of regard (blue, green).
  The video starts in the center of the spiral and time continues clockwise away from the center. The spiral pattern allows one to represent long time spans and supports a good overview, even on multiple recordings. The visualization can be created directly for the egocentric video and with gaze data in the global and local context of the video. Black scanlines are inserted for frames with missing gaze data.}
  \label{fig:teaser}
\end{teaserfigure}

\maketitle

\section{Introduction}

To capture gaze data in realistic, everyday situations, mobile eye tracking is deployed in many areas, such as behavioral research~\cite{Land2001}, pervasive computing~\cite{Bulling2011}, and augmented reality~\cite{Naspetti2016}.
It combines gaze data and egocentric video, two data sources already difficult to analyze on their own. Each recorded video consists of individual coordinate systems, movement, and timing. Hence, an analysis beyond descriptive statistics of gaze characteristics (e.g., average fixation duration) requires contextual information from all recordings. 
Especially for the analysis of scanpath sequences, relevant areas and objects have to be identified in each video, to this point often manually, rendering the definition of AOIs as one of the most time-consuming steps in the analysis process.

Interactive data visualization provides means to investigate recorded videos and their respective scanpaths under numerous aspects~\cite{Blascheck2017}.
Most common is the depiction of measures over time (e.g., pupil dilation, coordinates) that provide more details than aggregated descriptive statistics.
In the stimulus, gaze replays overlaid on the egocentric video can be investigated with most software packages shipped with the device. However, a comparison between multiple participants is difficult to achieve with gaze replays alone. Hence, most techniques to visualize mobile eye tracking data in a static overview require the definition of AOIs as a preprocessing step. With the information in which order people looked at specific objects or areas, sequential representations (e.g., scarf plots~\cite{Richardson2005}) help interpret data and compare between multiple participants~\cite{Kurzhals2014}. Overview and comparative visualizations that work without AOIs are sparse.
Although progress in machine learning techniques allows one to identify specific AOIs automatically~\cite{Wolf2018}, this typically requires pretrained categories for specific AOIs. An application to arbitrary eye tracking experiments is still limited to rather general object categories. For example, it is possible to detect when a person looked at different cars, but to identify when it was their own car requires external knowledge that has to be fed into the algorithm. Visualization can help communicate such human interpretations to the computer, making the data more accessible for analysts, and it helps interpret results from automatic processing steps.

We introduce a visualization approach for the overview, annotation, and comparison of numerous recordings from mobile eye tracking. This is achieved by creating a compact, static representation of video content incorporating gaze positions: the \emph{gaze spiral}. In this way, visual patterns such as dwells on an area, as well as changes between different areas of the stimulus are preserved. 
For example, Figure~\ref{fig:teaser} shows three types of spirals created under different conditions. The gaze spirals convey frequently occurring patterns of gaze behavior that can be visually identified.
%Although the link to the underlying stimulus is sometimes not obvious, those We argue that this link is not necessarily required to identify those previously mentioned cues.
The compact representation of the \emph{gaze spiral} allows to visualize even hour-long recordings and highlight results from queries without the need for horizontal or vertical scrolling.
Furthermore, multiple recordings can be displayed simultaneously for the interpretation of automatic comparison methods. 
Overall, we contribute a new technique to (1) visualize scanpaths from mobile eye tracking in a compact representation, (2) investigate visual behavior patterns by similarity queries and annotation, (3) and facilitate projection-based comparison of recordings. Our approach does not require AOIs and can be applied directly to recorded gaze data. Hence, we see this as a starting point for the analysis of data from mobile eye tracking experiments, to assess the quality of recordings, identify relevant scanpath patterns, and derive AOIs for established analysis techniques. %\vspace{-2ex}

\section{Related Work}\label{sec:related_work}
Relevant work with respect to this technique comprises comparable approaches that focus on the analysis of data from mobile eye trackers and more general visualization techniques for temporal data, especially video data.

\subsection{Mobile Eye Tracking Analysis}
Eye tracking data from mobile glasses can be analyzed without contextual information from an egocentric camera, for example, the detection of vigilance~\cite{Ji2002} or cognitive load~\cite{Rafiqi2015}.
We focus on techniques that include the video stimulus because they generally comprise more scenarios to investigate behavior such as spatial orientation~\cite{Kiefer2017} and activities in everyday situations~\cite{Land2001}.

As mentioned, with AOIs available, established means of statistical analysis can be applied and used for visualization. \citet{Blascheck2017} provide an overview of visualization methods based on AOIs. Techniques that were applied to mobile data are hierarchical techniques such as transition matrices and graph visualization~\cite{Blascheck2016}, as well as timeline-based techniques such as scarf plots~\cite{Richardson2005}.

Since the definition of AOIs is a complex problem by itself, some techniques have been proposed that visualize image content of the video directly while incorporating information about gaze positions. Techniques such as fixation-image charts~\cite{Kurzhals2016} visualized scanpaths from videos by thumbnail image sequences, extracted at the point of regard. Scanpath comparison techniques were applied based on image features~\cite{Kurzhals2015,Castner2020}, providing results comparable to AOI-based methods. This image-based approach was also proposed for the annotation of AOIs, supported by automatic processing of similar images~\cite{kurzhals2016a}. However, these methods did not focus on representing the context of the stimulus in its temporal order. Instead, by annotating AOIs in the visualization, the sequential scanpath was partially displayed by color-coded scarf plots. In contrast, we provide a visualization that supports comparisons between recordings without the need for annotation.

Multi-layered timelines with spectrograms for different feature intensities were proposed for long-term eye tracking recordings by~\citet{Kurzhals2020}. By querying selected time spans, similar sequences could be searched and were depicted by short thumbnail sequences of the recorded field of view with a gaze overlay. Although the authors investigated hour-long videos, they focused on the analysis of single videos. A comparison between recordings was not possible. With our approach, we aim to fulfill both requirements, by providing a compact representation of long video sequences and methods to compare between participants.

\subsection{Visualization of Temporal Data}

The visualization of video data can be separated into applications for analytical purposes and video-based graphics~\cite{Borgo2012}. We see our approach mainly on the analytical side, however, it is inspired by techniques for video artwork based on the slitscan technique, for instance, the \emph{Last Clock} application\footnote{https://gnomalab.es/last-clock-ipad/. Last checked on April 14, 2022.}, which presented the images resulting from video processing with a clock metaphor on iPads.

In the context of video analysis, this technique has been applied to create visual fingerprints that summarize long videos~\cite{Bezerra2006,Seitz2003}. 
\citet{Kurzhals2016} presented the idea to include gaze data to create slitscans for videos shown with a static remote eye tracking system. However, they did not apply the technique to mobile eye tracking data. \citet{Kurzhals2021} used slitscans as a visual aid to determine thresholds for the segmentation of video data. The visualization was represented as horizontal timelines. This layout impairs the comparison for long recordings because horizontal scrolling for individual timelines becomes necessary once the number of timesteps exceeds the horizontal screen resolution. 
\citet{Burch2018} suggested applying dimensionality reduction on subsequences of scanpaths from static images, but the visualization of the results was restricted to scatterplots, impairing a direct interpretation of the results.
\citet{Weber2001} argued that spiral visualization of time-series data scales well with time and helps identify periodic and recurring patterns.
A linear representation either needs line wraps, excessive zooming, or scroll interactions to be displayed on the screen. Compared to that, our approach produces a continuous scanpath representation without line wraps. Zooming still might be required to fit the entire sequence on the screen, but to lesser degree than linear representations.

\section{Technique}\label{sec:technique}

\begin{figure}[t]
    \centering
    \includegraphics[width=0.9\textwidth]{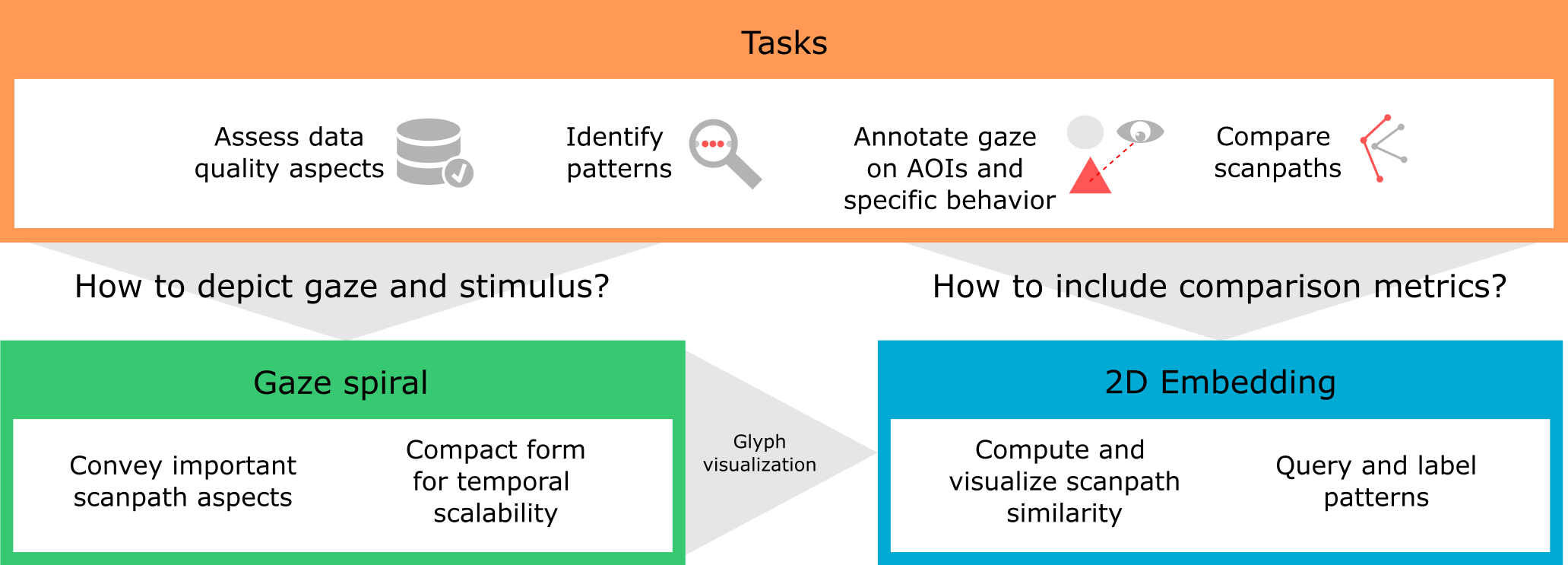}
    \caption{Gaze spirals serve as compact representation of visual context and gaze properties of a scanpath. When investigating multiple recordings, they can be applied as glyphs in a 2D embedding, based on scanpath comparison metrics. The presented approach aims to support four common tasks for eye tracking analysis: the visual assessment of data quality, identification of patterns, annotation of AOIs and behavior, and the comparison of scanpaths from different recordings.%\vspace{-2ex}
    }
    \label{fig:technique_overview}
\end{figure}

We designed a visualization technique for data from mobile eye trackers in unrestricted scenarios. Based on the literature~\cite{Kurzhals2017} and our own experience with eye tracking experiments, we identified four tasks (Figure~\ref{fig:technique_overview}) we address in this context: 

    \paragraph{Data Quality Assessment} 
    Data quality has numerous aspects~\cite{Holmqvist2012}; we help assess quality in terms of data loss because it is a prerequisite for most analyses.
    Since eye tracking is prone to issues with the detection of gaze samples, data will be missing for a multitude of reasons. Important is that experimenters are aware of missing data and can decide whether a recording is sufficient to be included in the analysis. This is often determined by defining a threshold for the acceptable percentage of valid samples. Investigating the temporal distribution of missing data is helpful~\cite{Schulz2015}, as short periods of missing samples might be less problematic than long periods without gaze detection during important events.
    
    \paragraph{Pattern Identification}
    Mobile eye tracking data consists of multiple data sources, i.e., recorded eye movements and a stimulus video that provides context. 
    In this work, we assume that patterns are manually identified by the user and that the visualization aims to support this process.
    Patterns can be identified in both, e.g., horizontal scanpath patterns during reading~\cite{Kunze2013}, or a specific sequence of how AOIs are investigated. Our main focus is on experiments in unrestricted environments where the context of the stimulus is necessary to identify behavioral patterns and visual strategies people apply to solve a task.
    
    \paragraph{Annotation}
    AOIs are the prerequisite for the majority of statistical and visual analysis methods in this mobile context~\cite{Blascheck2017}. If gaze samples or fixations can be assigned to a specific AOI, distributions can be calculated to identify frequency and duration of gaze on each AOI. Similarly, behavior related to eye movement (e.g., visual search) often needs to be identified and annotated. In many scenarios, especially in in-the-wild studies, a fully automated solution is not possible and manual annotation is necessary. 
    
    \paragraph{Scanpath Comparison}
    Scanpath comparison~\cite{Anderson2015} is applied for the identification of common visual strategies (e.g., how people read metro maps~\cite{Netzel2017})  or the grouping of levels of expertise (e.g., for chess players~\cite{Reingold2011}). For mobile eye tracking data, this is typically achieved with string-comparison methods of annotated data. These methods provide pairwise similarity values that can be clustered for a larger number of recordings, resulting in groups of similar scanpaths.

\begin{figure}[t]
    \centering
    \includegraphics[width=0.9\textwidth]{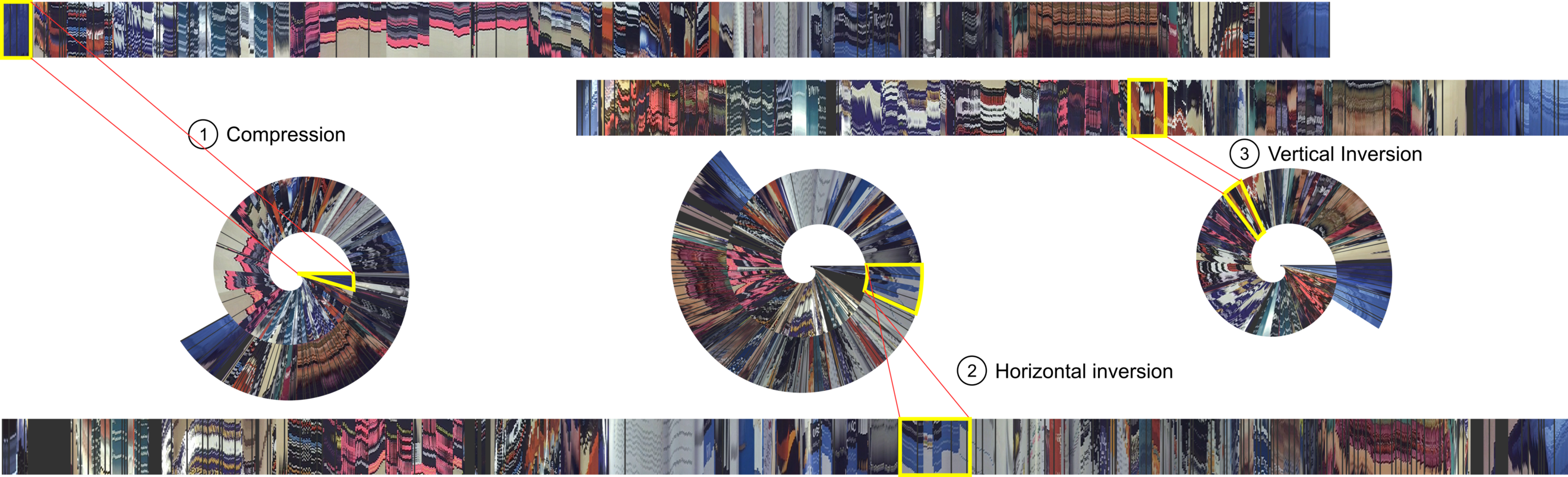}
    \caption{Comparison of linear and spiral gaze-guided slitscans for three recordings. (1) At the center of the spiral, the compression of visual content is the strongest, which decreases with later timesteps. (2) The linear representation is displayed with a reading direction from left to right, a clock-wise spiral inverts the respective patterns horizontally. (3) If the rotation of the scanlines is continuous, the patterns in the upper part of the spiral are vertically inverted. In general, the spiral uses display space more efficiently than the linear slitscan.
    %\vspace{-2ex}
    }
    \label{fig:linear_spiral}
\end{figure}

To address all these tasks together, existing techniques are not sufficient, either because they require annotated data (e.g., scarf plots, comparison metrics), do not preserve the sequential order of a scanpath (e.g., gaze thumbnails~\cite{kurzhals2016b}), or are just not well suited for long sequences. Hence, we designed a compact representation of slitscans that conveys important scanpath information, particularly, what visual context was investigated over time by combining information about the point of regard and the stimulus. This single representation of a scanpath---the \textit{gaze spiral} (Section~\ref{sec:gaze_spiral})---can serve as a compact glyph that conveys important scanpath aspects and is further applied to compare numerous recordings.
We discuss the parameter space for gaze spirals and how individual parameters influence the rendered results (Section~\ref{sec:parameters}).
Important patterns indicating gaze on AOIs can be queried and annotated directly in the visualization (Section~\ref{sec:query}).
For comparison, we create a 2D embedding of all scanpaths based on established comparison metrics (Section~\ref{sec:2d-embedding}). The result shows gaze spirals in a spatial layout where similar scanpaths are in spatial proximity and visual patterns, mainly referring to gaze on AOIs, can be investigated, queried, and annotated in the visualization. 

\subsection{Gaze Spiral}\label{sec:gaze_spiral}
A slitscan is generated by placing a scanline in a video, extract pixels along the scanline over time, and render the scanlines into a static overview. 
The technique is not limited in terms of location and orientation of the scanline~\cite{Tang2008}. For gaze-adaptive slitscans~\cite{Kurzhals2016,Koch2018}, the location of the scanline is adjusted, according to the current point of regard. 
For the representation, horizontal timelines have the advantage that they represent linear time in the most common way~\cite{Aigner2011}. 
Considering the scalability of temporal extent, individual timesteps are bound to the screen resolution, which means that for long sequences, the visualization either has to be scrolled, making it hard to compare numerous timelines, or zoomed out, which results in a loss of visual information. Hence, we suggest applying a spiral pattern to represent slitscans (Figure~\ref{fig:linear_spiral}).
Different patterns, such as a space-filling curve~\cite{Dafner2000}, would show the data with a comparable density. 

We decided for the spiral because it represents time with a clock metaphor, which is familiar to most people. Spiral data visualization can further help detect recurring events, which is often the case in eye tracking data, e.g., when people are repeatedly switching attention between different areas of interest. Further, the radial representation is often perceived as aesthetic, which renders the visualization as a good method for the illustration and communication of the data.
We argue that the slitscan representation comes with a good trade-off between compactness and interpretability. 
In principle, other types of representations could be wrapped into a spiral form, but they have their own set of problems.
For example, gaze stripes~\cite{Kurzhals2015} are based on image thumbnails, thus providing more context to the underlying stimulus than scanlines. However, this property makes gaze stripes also impractical for longer recordings. In contrast, scarf plots are compact but assume the existence of AOI definitions, which are difficult to provide in free-viewing contexts.

For data visualization, the Archimedes' spiral is the most commonly applied spiral form~\cite{Weber2001}. 
It reads in polar coordinates:
%, as it provides equidistant traces along which the scanlines can be placed:
\begin{align*}
    \phi(t) &= t^k \\
    r(t) & = a \cdot \phi(t)/2\pi
\end{align*}
Here, time $t$ is a unitless quantity and increases with a fixed step size. It can be derived from physical time (recording time or frame number) by applying appropriate scale and bias operations. We assume that the data starts at $t=0$.

The Cartesian coordinates ${x(t) = r(t) \cdot \cos(\phi(t))}$ and ${y(t) = r(t) \cdot \sin(\phi(t))}$ define the position of the spiral baseline.
Finally, the gaze spiral is generated by placing the scanlines orthogonal onto the spiral's baseline, pointing outward from the perspective of the baseline.
The parameter $a$ is a constant for the distance between spiral arms and the rotation angle between timesteps can be influenced by the exponent $k$.

\subsection{Parameter Space}\label{sec:parameters}

For the design of gaze spirals, a set of parameters can be adjusted to influence the outcome of the rendering process.

\begin{figure}
    \centering
    \begin{subfigure}[t]{.45\linewidth}
        \centering
        \includegraphics[width=\linewidth]{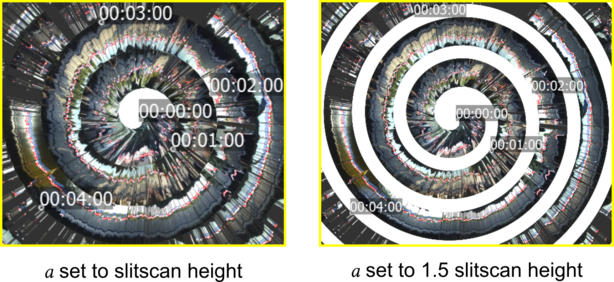}
        \caption{Distance $a$}
        \label{fig:distance_comparison}
        \end{subfigure}
        \hfill
        \begin{subfigure}[t]{.45\linewidth}
        \centering
        \includegraphics[width=\linewidth]{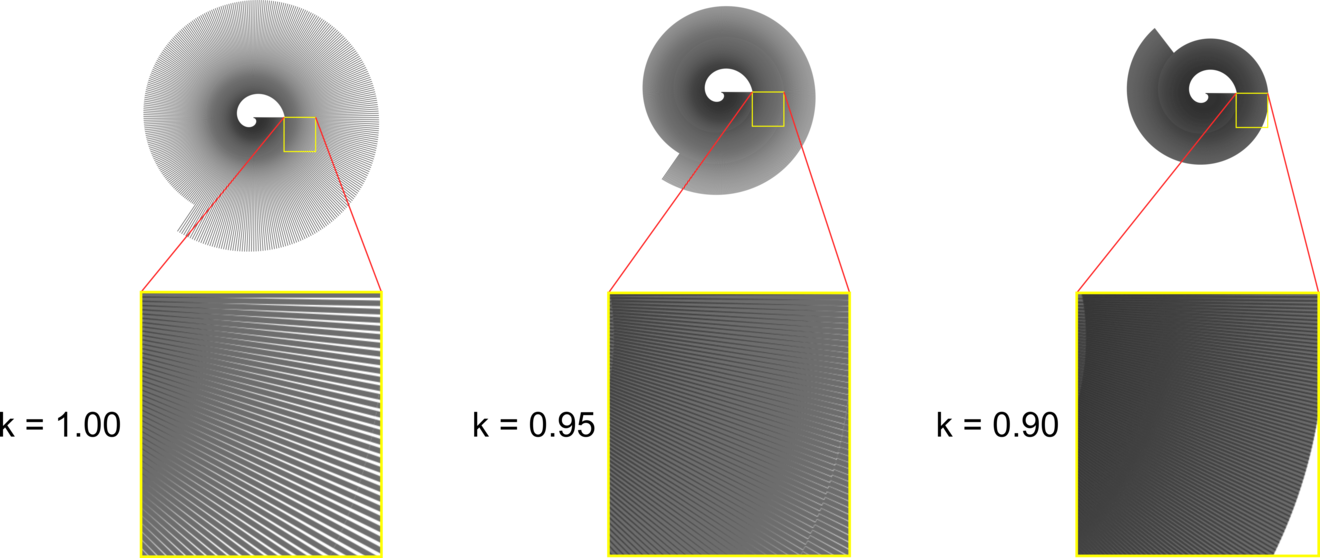}
        \caption{Angle $k$}
        \label{fig:angle}
    \end{subfigure}\vspace{2ex}
    \begin{subfigure}[t]{0.9\linewidth}
        \centering
        \includegraphics[width=\linewidth]{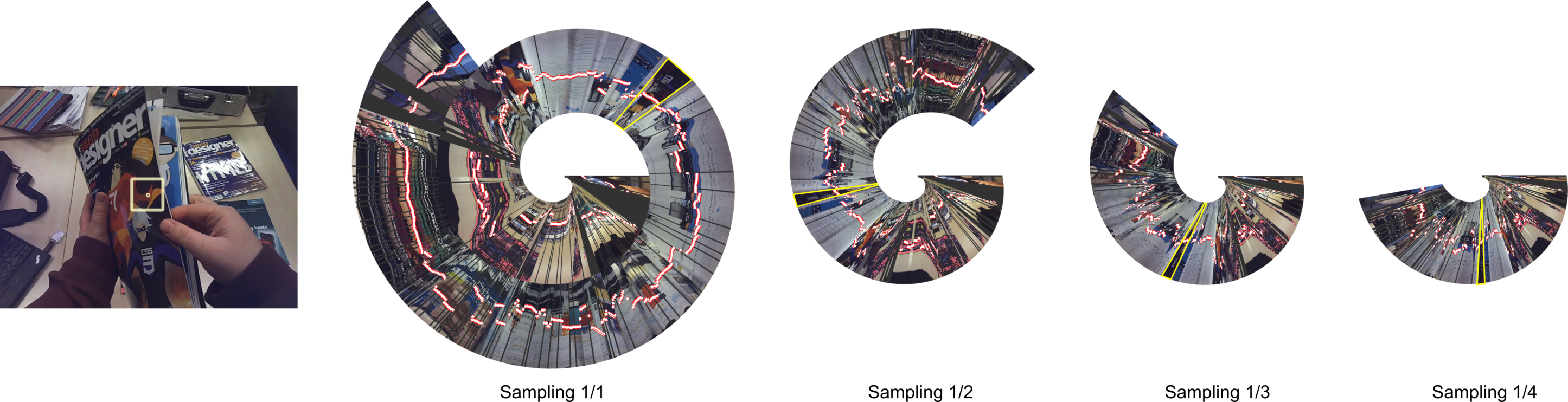}
        \caption{Sampling of frames}
        \label{fig:sampling}
    \end{subfigure}
    \caption{(a) Blank space in the spiral is adjusted by $a$. It helps identify timesteps of a label, at the cost of the compactness of the visualization. (b) Influence of $k$ on the resulting visualization.
    Decreasing the value results in a smaller distance between scanlines. Consequently, the length of the spiral is also decreased. (c) Uniform reduction of samples decreases the size of the spiral significantly. Patterns resulting from fixations diminish, the more frames are skipped. The presented example shows a fixation with a duration of approximately 500\,ms (yellow highlight). %\vspace{-2ex}
    }
    \label{fig:parameter_space}
\end{figure}

\paragraph{Level of Detail}
First, the data can be represented with different levels of detail, mainly by focusing either on the egocentric video, or the underlying gaze data (Figure~\ref{fig:teaser}). Focusing on the egocentric video, a vertical scanline in the center of the video can be applied. 
Such data is often used as an approximation for visual attention, given the assumption that people mainly focus on the center of their field of view. However, as Figure~\ref{fig:teaser} shows, this is not always the case and including eye tracking data provides further details about the viewing behavior.
With global gaze-adaptive slitscans, the $x$-position of the point of regard is applied to position the scanline. The $y$-position can be indicated with a visual marker, in the example rendered as white dots with a red border. Due to the rapid nature of eye movements, the smoothness of patterns is interrupted, similar to slitscans resulting from edited video material. 
Local gaze-adaptive slitscans focus on a small area around the point of regard. While this provides the most details on a fixation level, the resulting slitscan is visually the least smooth, as it is disrupted by horizontal and vertical eye movements. 

\paragraph{Distance}
By increasing $a$, additional blank space between the arms of the spiral can be created (Figure~\ref{fig:distance_comparison}). While this impairs the compactness of the spiral, it also provides new space to render additional information into the visualization. For instance, annotations or related eye tracking metrics could be integrated by color codings~\cite{Koch2018}.
We propose setting $a =1.2$ (where the slitscan height is assumed to be 1) and placing a fixation color encoding into the empty space.
For that, prior to rendering the gaze spiral, we extract fixations from the gaze and video data using the method described by \citet{SteilHB18}.
For the purpose of applying the spirals as small multiples, we suggest favoring the compact representation without labels or color coding.

\paragraph{Angle} The angular sampling of the visualization determines how dense consecutive scanlines will be placed in the visualization. The angle also influences the total size of the spiral. We control the angle mapping via the parameter $k$.
The pattern at $k=1$ corresponds to equidistant samples in time (Figure~\ref{fig:angle}).
As one can see, the distance between sample points on the spiral increases quickly.
%, which leads to blank space between the rendered scanlines of the slitscans.
%To avoid visual artifacts resulting from blank spaces between slices when the radius increases, we slightly decrease $k$ to create slitscans without blank spaces in the spiral. 
In general, setting $k < 1$ compresses the spiral, whereas setting $k > 1$ spreads the spiral. We typically use $k=1$ (for equidistant sampling) or values slightly less than one for moderate reduction of the angular speed toward the outward parts of the spiral (e.g., $k = 0.9$).

%\paragraph{Scanline Width}
%As discussed, blank space resulting from an increased radius can be reduced by adjusting $k$. Further, the space can also be filled by increasing the width of the scanline. While this increases the amount of displayed visual information from one timestep, it will also overdraw previous timesteps, leading in the worst case to a shift between the image content and its corresponding temporal location on the spiral. 

\paragraph{Scanline Orientation} For the described technique, we applied a vertical scanline. However, as \citet{Tang2008} have shown, it is possible to take arbitrary orientations and lengths for the scanlines to focus on specific motion and locations. We focus on the current point of regard, which determines the location of the scanline. Considering the orientation, we applied vertical scanlines for consistency between the different renderings. However, with respect to gaze behavior, it might be beneficial to orient the scanline according to the angle between consecutive fixations.

\paragraph{Sampling}
The sampling of the data can be reduced to decrease the size of the resulting spirals. With respect to eye tracking, the sampling of this data source is adjusted to the sampling rate of the video. Hence, a video with 25\,fps results in 40\,ms duration per frame. With an average fixation duration of 200--300\,ms~\cite{Rayner1998} for information processing, we do not recommend reducing the sampling of the data below one fourth of the original samples, since it could result in missing static patterns typical for fixations. Figure ~\ref{fig:sampling} displays an example of how the visualization changes when the samples are reduced uniformly. Future work could consider adaptive sampling techniques to represent only important data and to reduce redundant information.

\subsection{Query and Annotation}\label{sec:query}

So far, the gaze spiral is static but user interaction is necessary for performing annotation tasks. 
The annotation system is based on four aspects: inspecting fixations;  selection of fixations; annotation of selections; and accepting recommendations.
Thumbnail images associated to each fixation in the user selection support the visual inspection of fixations.
This selection is adjustable, thus the user can skim through the whole sequence fixation-by-fixation.
This selection can also be expanded or shortened upon user request.
The actual annotation is, when initiated by the user, always performed on the current selection of fixations.
Annotation also automatically triggers a query mechanism to find candidate fixations that are similar in terms of image content. 
These candidate fixations are placed as thumbnails along the spiral.
The user now can continue the annotation process from those candidate fixations by a mouse click on the corresponding thumbnail image, which anchors the user selection on that specific fixation.

\subsection{2D Embedding}\label{sec:2d-embedding}

After discussing how a single spiral is created, this section considers how to compare different spirals and respective recordings.
Scanpaths can be compared with string-based sequence alignment algorithms, like the Levenshtein distance \cite{Levenshtein1966} or the Smith-Waterman algorithm \cite{Smith1981}.
String-based representations are usually obtained from scanpaths by mapping each fixation to a symbol that corresponds to a specific region in the stimulus.
Instead of relying on AOIs, we use a feature-based representation of fixations, which has been proposed before by \citet{Kurzhals2015} and \citet{Castner2020} to perform sequence alignment between scanpaths.
For the Levenshtein and Smith-Waterman algorithms, the cost is defined by the cosine distance.
The gap cost for both algorithms needs to be defined prior to alignment.
In our following discussions, we define the gap cost for both algorithms to be $0.5$ for feature-based sequences and to be one for string-based sequences.
Additionally, we use Dynamic Time Warping~\cite{BerndtC94} to perform alignment without the need of specifying gap costs.

\begin{figure*}[t]
    \centering
    \includegraphics[width=0.9\linewidth]{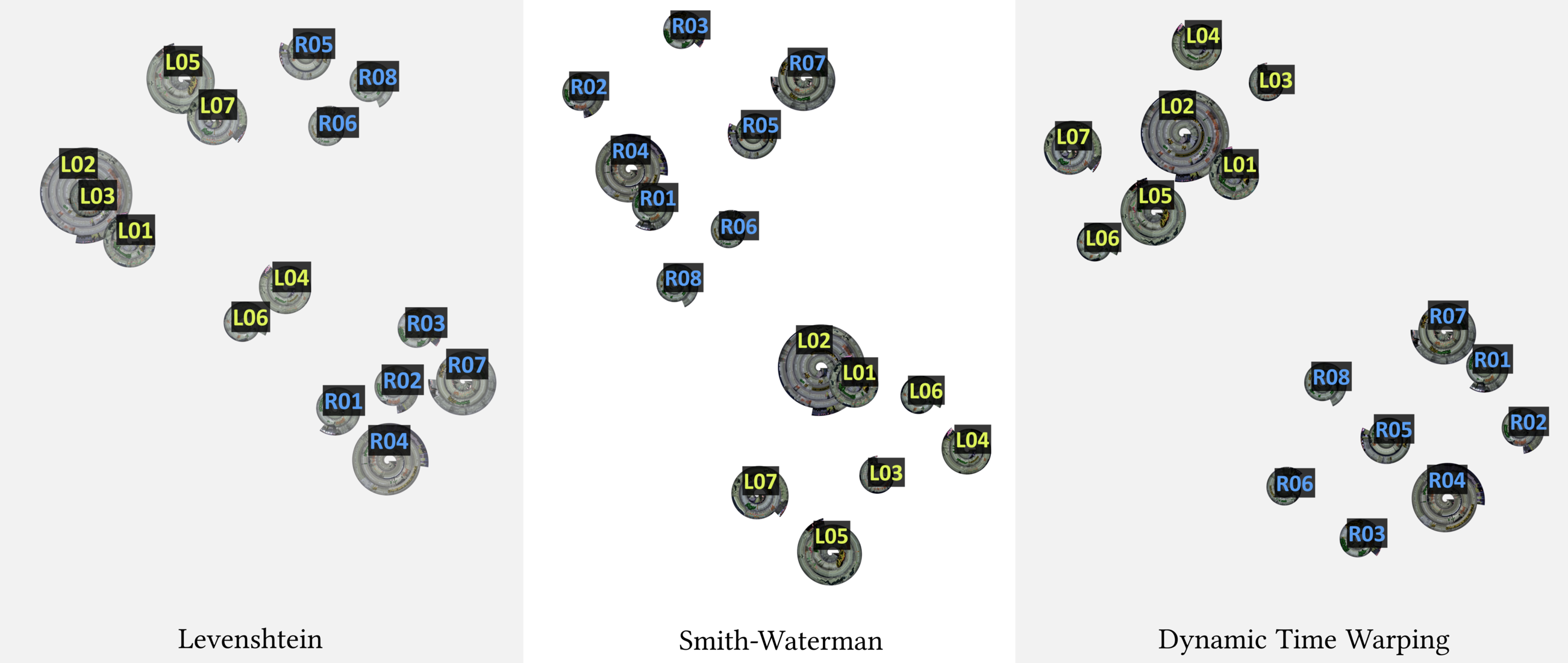}
    \caption{2D embeddings of UMAP projection algorithm with three different metrics applied to 15 scanpaths. Each scanpath is represented by a gaze spiral, and its screen position is defined by the 2D embedding.}
    \label{fig:embedding-metric-c}
\end{figure*}

Hereafter, we present a visual approach to support the interpretation of metric outputs.
First, the distances between all pairs of sequences is computed with the target metric.
The resulting matrix of pairwise distances is used to calculate a 2D embedding of scanpaths using dimensionality reduction. 
Notice that the previously described preprocessing is identical to that performed before clustering.
Each scanpath is visually represented by its corresponding gaze spiral, and its center position on the screen is defined by the 2D embedding.
Figure \ref{fig:embedding-metric-c} shows an embedding of 15 scanpaths (see Section~\ref{sec:dataset}).
The 2D embedding is obtained with Uniform Manifold Approximation and Projection (UMAP)~\cite{Mcinnes2018}, which is a state-of-the-art technique for dimensionality reduction.
As shown in the Figure \ref{fig:embedding-metric-c}, depending on the applied metric, the embedding results in different local clusters.
Consequently, it is necessary to investigate individual clusters and interpret why they were rated as similar by the respective algorithm. This task can be performed directly in the visualization with the gaze spirals. An example is discussed in Section~\ref{sec:comparison}.

\section{Comparison with established Methods}\label{sec:comparison}

The purpose of the following analysis is to evaluate how a 2D embedding of gaze spirals can be interpreted and how the approach compares to established methods that rely on annotated data, i.e., Hierarchical Cluster Analysis (HCA) on string-based AOI sequences.

\subsection{Dataset and Annotation Scheme}\label{sec:dataset}

\begin{figure*}[b]
    \centering
    \includegraphics[width=0.9\linewidth]{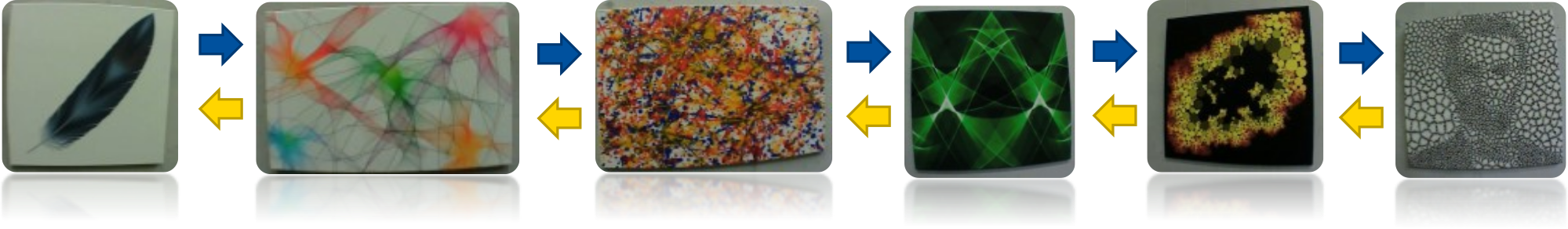}
    \caption{The art collection scene consists of six different paintings of algorithmic artwork. Each participant was asked to attend the paintings in two specific orders as indicated by the blue and yellow arrows.}
    \label{fig:dataset}
\end{figure*}

To showcase our approach, we recorded a dataset that emulates a scenario for the application of mobile eye tracking, the visit of an art gallery.
The scenario is mainly defined by a set of AOIs, i.e., different paintings and the description text next to each piece. Visitors walk through the gallery, look at the paintings, read descriptions, and switch their visual attention in different order and duration between AOIs. Hence, it could help investigate how people view the paintings, for example, to rearrange some parts of the gallery. Further, there might be some groups in the data with different viewing behavior, for example, children or elder people. Note that the data was recorded for illustration purposes and does not claim any scientific significance about the interpretation of the presented stimuli. 
In the following, we will show how our approach could be applied to answer such questions and compare how this would be achieved with established methods.

We asked 8 participants to walk through an art gallery at our institute, investigate the artwork, and read the text. This task was performed twice, once in order from left to right (R) and once from right to left (L), simulating a first and a second impression of the work. The starting order was altered between participants. 
Data was recorded with Pupil Invisible glasses. An offset correction was performed at reading distance to text descriptions.
This resulted in 16 scanpaths, one was removed due to recording issues.
The scanpaths contain two induced overall patterns (L and R), and individual differences between participants looking at the AOIs.  

The gallery consists of six paintings of algorithmic art (Figure~\ref{fig:dataset}), each defined as an individual AOI.
Additionally, we define one AOI that encompasses the description texts of all paintings and one for all remaining areas. Fixations on AOIs were annotated with our approach and serve as the baseline we compare against.

\subsection{Comparison between Hierarchical Cluster Analysis and 2D Embeddings}

\begin{figure*}[t]
\centering
\begin{subfigure}{.49\textwidth}
  \centering
  \includegraphics[height=1\linewidth]{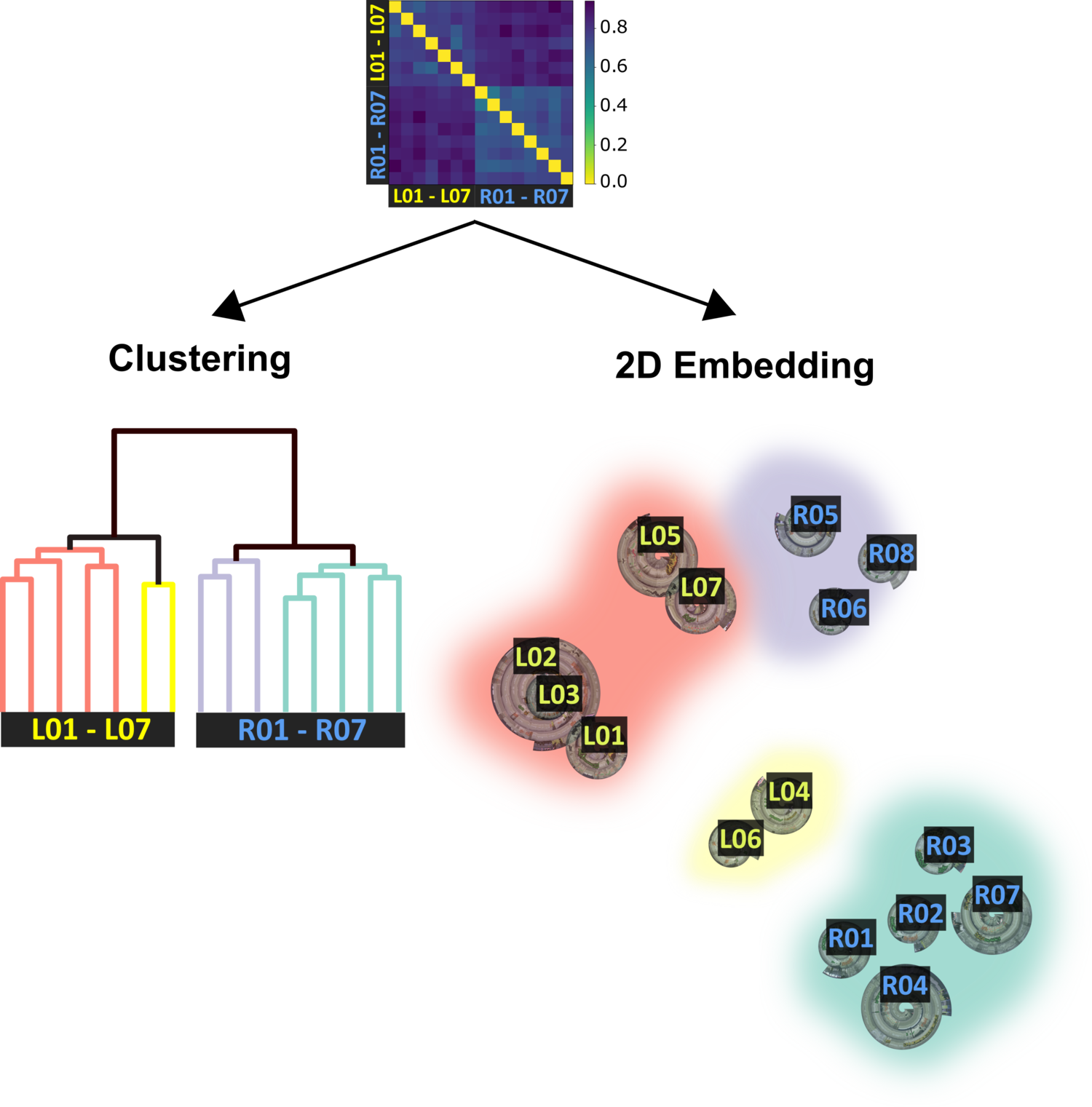}
  \caption{Levenshtein feature-based alignment.}
  \label{fig:sub1}
\end{subfigure} \hfill
\begin{subfigure}{.49\textwidth}
  \centering
  \includegraphics[height=1\linewidth]{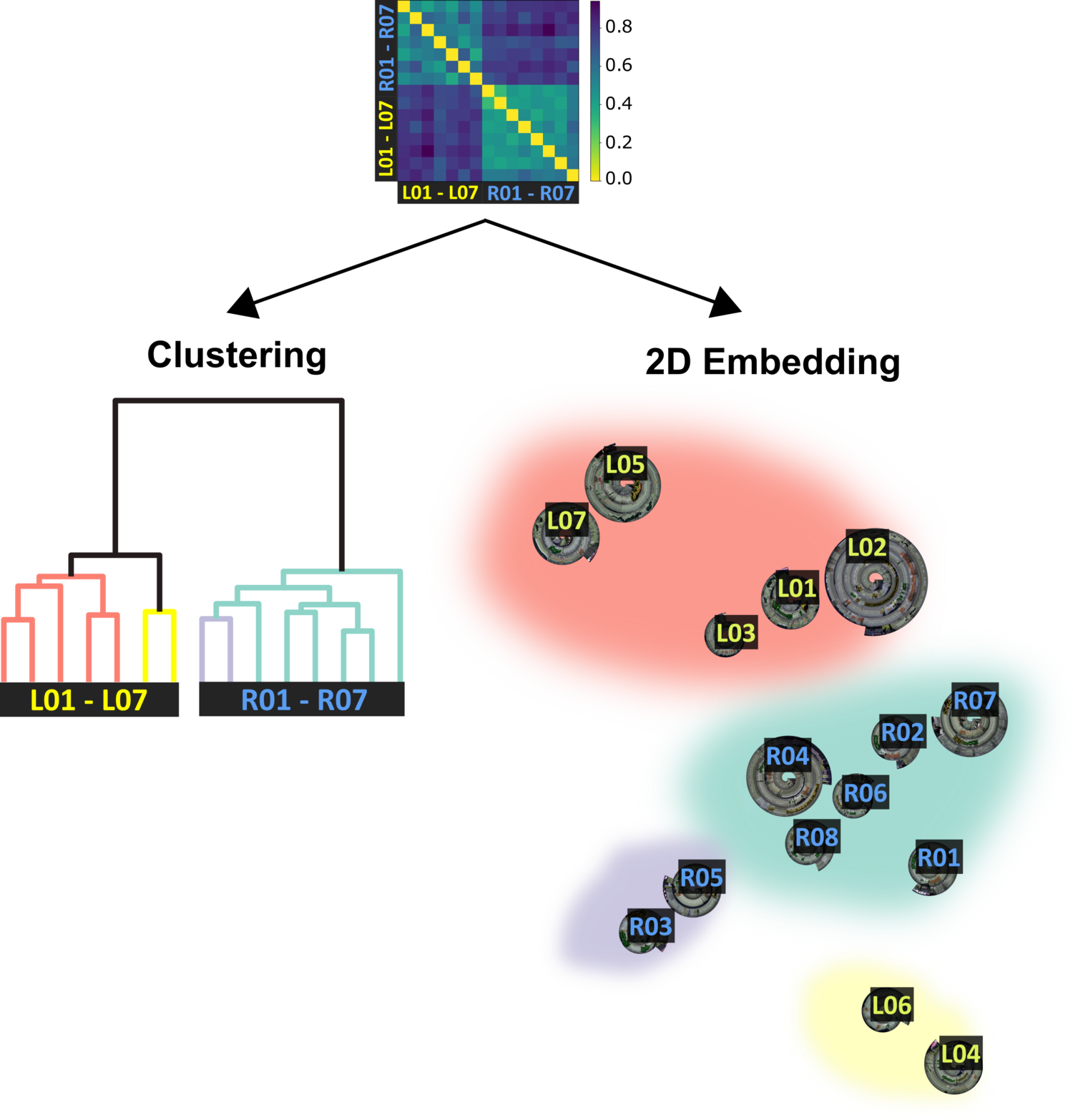}
  \caption{Levenshtein AOI-based alignment.}
  \label{fig:sub2}
\end{subfigure}
\caption{Comparison between 2D embedding and Hierarchical Cluster Analysis (HCA) based on two different types of sequence alignments (a) and (b).
Each alignment between the 14 sequences results in a pairwise distance matrix from which the clustering and 2D embedding are generated.
Both outputs show that clustering and 2D embedding lead to comparable results.}
\label{fig:clustering-vs-embedding} 
\end{figure*}

First, we investigate the pairwise distances between all sequences in the dataset under consideration of the different alignment methods (see Section~\ref{sec:2d-embedding}).
Taking into account the correlation between AOI-based (A) and feature-based sequence (F) alignment, we calculated the Pearson correlation between distance matrices. We get the correlations ${\kappa(\text{Levenshtein-A}, \text{Levenshtein-F}) = 0.939}$, ${\kappa(\text{Smith-A}, \text{Smith-F}) = 0.901}$, and ${\kappa(\text{DTW-A}, \text{DTW-F}) = 0.904}$.
The highest correlation between feature and AOI alignment can be observed with the Levenshtein distance.
This shows that, independent of the applied alignment method, the calculated similarities based on image features, which can be performed directly on the recorded data, highly correlate with annotated results.

Second, we want to compare how clusters derived from annotated data compare to cluster results in the 2D embedding.
Figure~\ref{fig:clustering-vs-embedding} illustrates the visual comparison of these results.
There is strong correspondence between the HCA result represented by the dendrogram and the 2D embedding, in particular on the lowest dendrogram level (colored groupings in red, yellow, purple, green).
The induced pattern between different viewing directions (L and R) is easy to identify at the highest level of the cluster hierarchies.
Going up the dendrogram hierarchy, we observe some larger deviations to the 2D embedding.
For example, in (b) the red and yellow groups belong to the same high-level cluster according to the dendrogram, but are rather far apart in the 2D embedding.
It should be noted that preserving distances in the 2D embedding cannot be guaranteed in UMAP at any granularity and depends strongly on the number of neighbors factor.
In general, a higher neighbors factor better preserves the global structure of the distances at the cost of fewer local clusters.
A neighbors factor = $3$ and higher results in a clear separation between L01--L07 and R01--R07, but misses the local clusters that can be observed when neighbors factor = $2$. Next, we want to investigate how clusters differ.

\subsection{Detailed Cluster Analysis}

\begin{figure*}[t]
    \centering
    \includegraphics[width=0.9\linewidth]{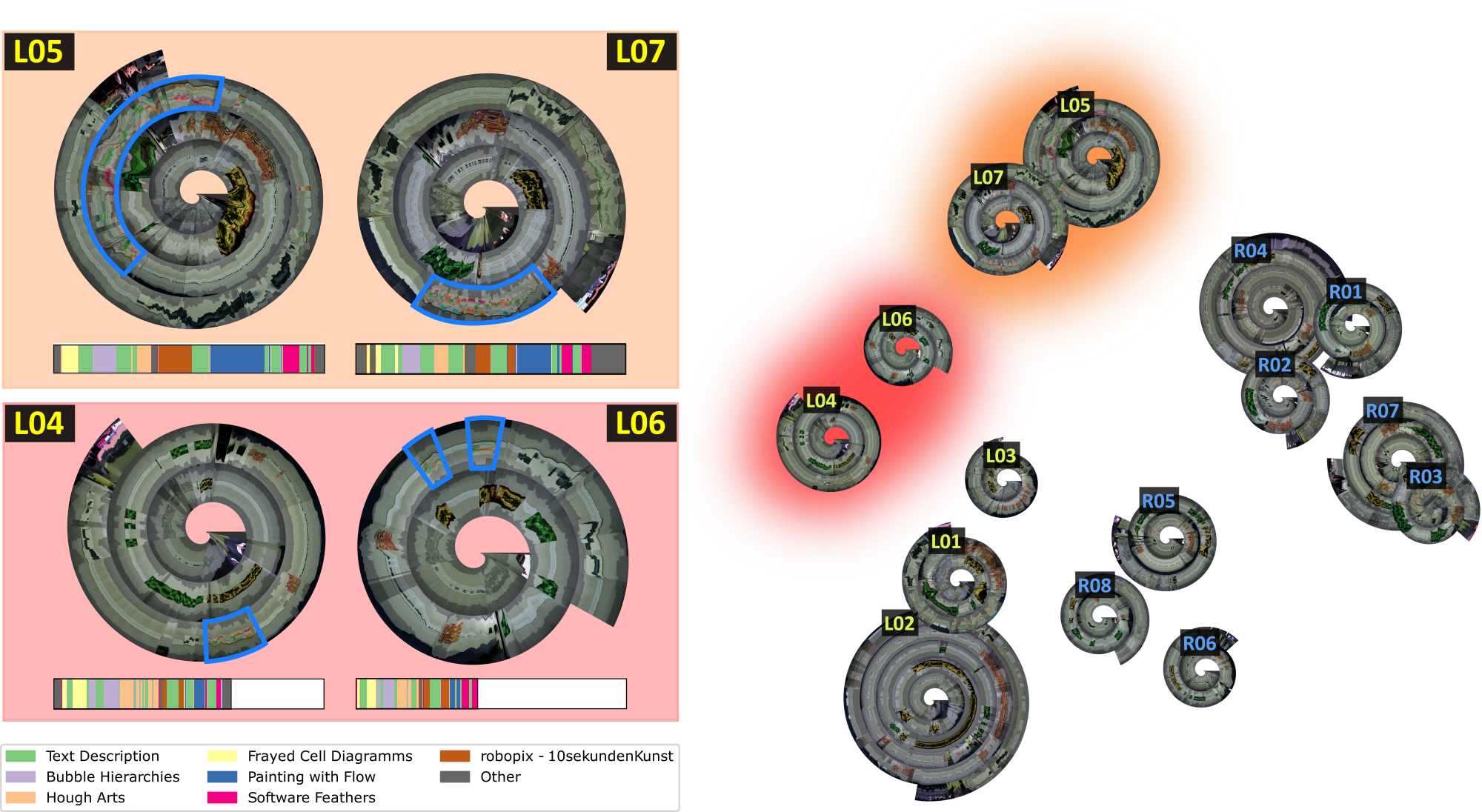}
    \caption{Right: 2D embedding of 15 scanpaths (L01--L07; R01--R08) based on Levenshtein distance. Left: Detailed analysis of between the clusters L05, L07 and L04, L06. Scarf plots of hand-labeled activities are shown beneath each gaze spiral. The spirals on the left side are normalized in size for better readability. %\vspace{-4ex}
    }
    \label{fig:embedding-metric-comparision}
\end{figure*}

Quantitative analysis of scanpaths based on Hierarchical Cluster Analysis fails to provide sufficient details on \textit{why} a particular grouping emerged.
The 2D embedding of scanpaths, visually represented with gaze spirals, aids such interpretation on a qualitative level.
Figure~\ref{fig:embedding-metric-comparision} provides a more detailed view into the 2D embedding obtained from Levenshtein distance. 
If we focus on individual clusters between scanpaths, we can investigate, for example, 
why the sequences L04, L06 and L05, L07 emerged as distinct groups in the embedding. 
L05 and L07 show a distinctive pattern that is less prominent in the scanpaths L04 and L06 (highlighted in blue). 
If AOIs are available, we can visualize the scanpath with scarf plots.
The scarf plots confirm that the dark blue AOI is more prominent in the upper cluster. Hence, scanpaths L05 and L07 attended the AOI \textit{Painting with Flow} longer than L04 and L06.
Similar investigations could be performed to describe the characteristics of individual scanpath groups, for example, to classify behavior.

\section{Examples}

\begin{figure*}[t]
    \centering
    \includegraphics[width=0.9\linewidth]{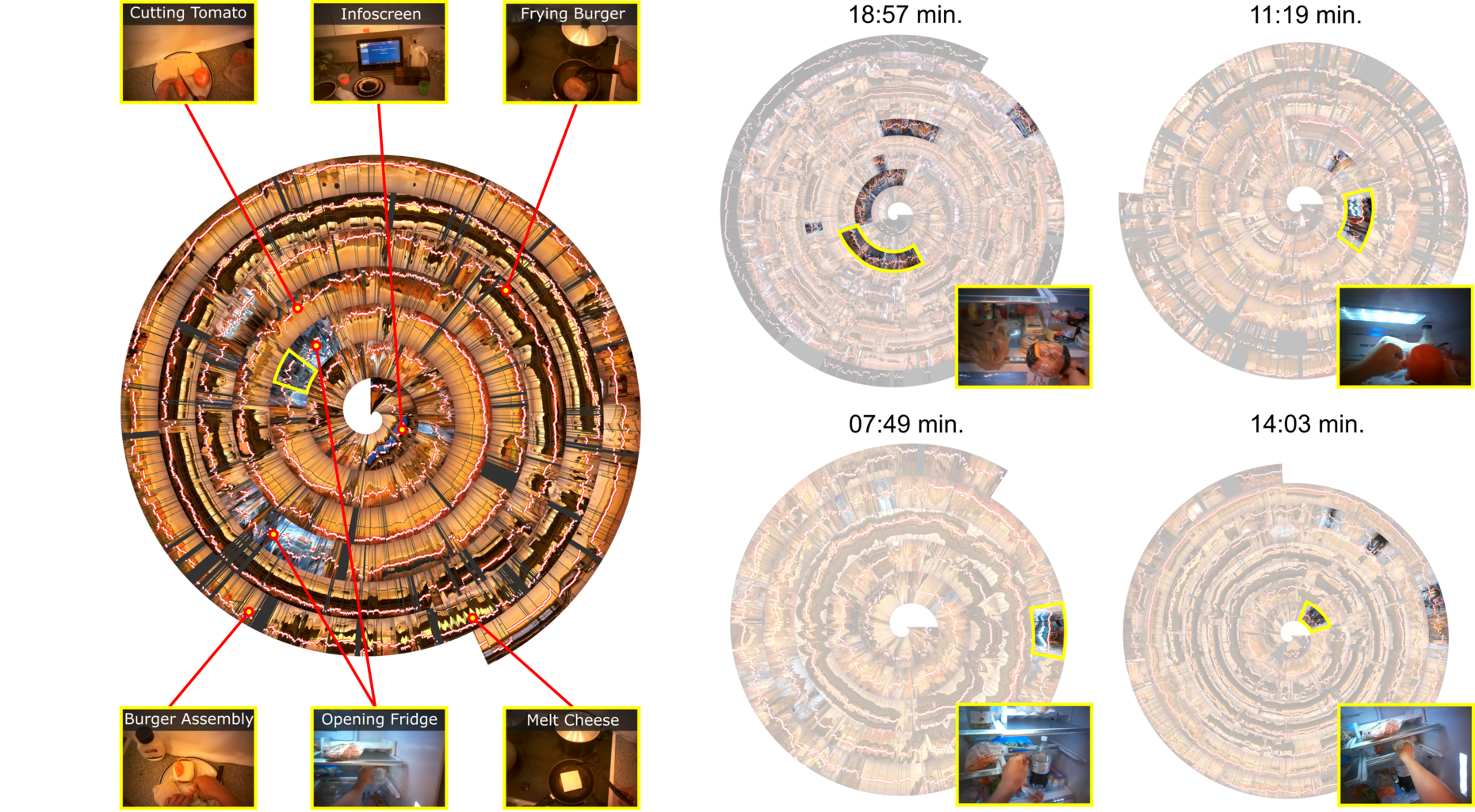}
    \caption{Example of gaze spirals for five sequences of the Extended GTEA Gaze+ dataset. In all sequences, the participants were asked to prepare a hamburger dish. Left: Annotated spiral with occurring events during the recording. Right: Results of an example query to find out when people opened the fridge. Time spans similar to the yellow framed slitscan pattern in the left spiral are visually highlighted in the other four spirals. %\vspace{-2ex}
    }
    \label{fig:multi_recording_comparison}
\end{figure*}

\begin{figure}[t]
    \centering
    \includegraphics[width=0.5\linewidth]{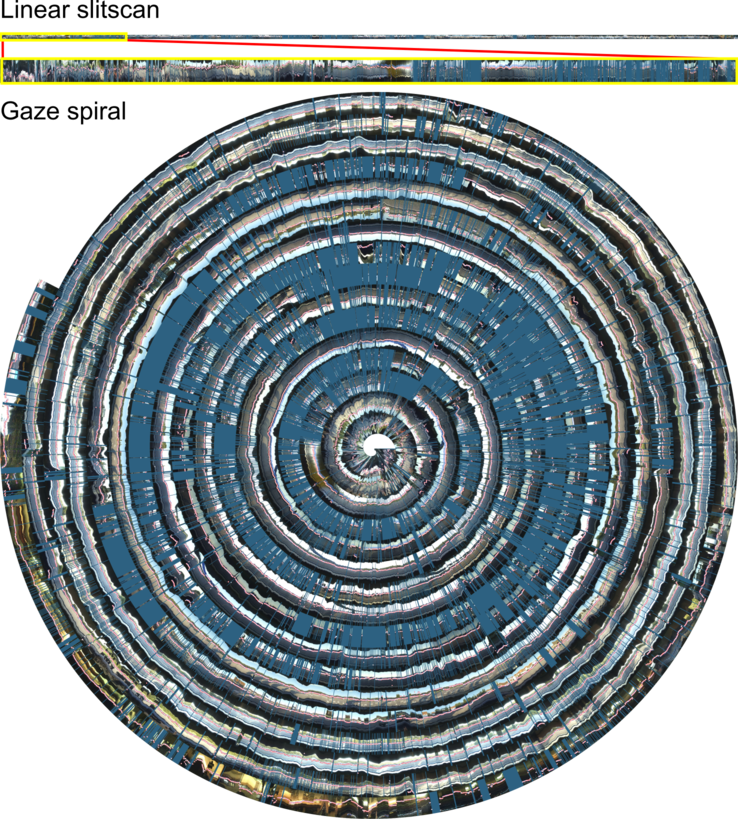}
    \caption{A linear slitscan and gaze spiral showing a video of a car ride (29:08 min). To display the 367,122 timesteps, we had to rescale the linear slitscan to 5\% of its original width. Even with this drastic reduction, only a short time span can be investigated when scaled to a height comparable to the gaze spiral. %\vspace{-3ex}
    }
    \label{fig:scalability}
\end{figure}

The presented examples (Figures \ref{fig:teaser}, \ref{fig:linear_spiral}, and \ref{fig:parameter_space}) comprise three videos of a dataset presented by~\citet{kurzhals2016a}. All videos show a person looking at four magazines, picking one up, skimming through it, and putting it back on the table. All videos are below one minute recording time.
For Figures~\ref{fig:distance_comparison} and \ref{fig:scalability}, we used a video recorded of a car ride~\cite{Kurzhals2021}.

To further demonstrate our approach, we investigated recordings from the Extended GTEA Gaze+ dataset for activity recognition in egocentric video~\cite{Li2018}.
This dataset contains 28~hours of content from 86 unique sessions of 32 participants performing several cooking activities.
Figure~\ref{fig:multi_recording_comparison} shows the gaze spirals of five different participants preparing a hamburger dish.
The duration of the shown sequences is between 8 and 19 minutes.
We used $k = 0.8$ for rendering and retained all samples to generate the spirals with high level of detail.
The left side of Figure~\ref{fig:multi_recording_comparison} depicts events during the preparation of a hamburger dish, e.g., cutting a tomato, looking at an information screen, or frying the burger.
Reoccurring events such as opening the refrigerator can be identified in the spiral as blueish patterns (highlighted by yellow borders 1--4).
Considering sequences from multiple subjects, one could be interested in finding similar events to compare how participants organized the preparation of the ingredients.
Based on the previously described query mechanism (Section~\ref{sec:query}), we selected the slitscan pattern in the left spiral of Figure~\ref{fig:multi_recording_comparison} (yellow framed segment) to query events in sequences from four different participants.
The right side of Figure~\ref{fig:multi_recording_comparison} shows the result of this query. Investigating the thumbnail images confirms that there is indeed a close similarity of all found time spans to the event of interest.
Most importantly, we can identify the results in the overview visualization immediately and the slitscan further helps interpret if a highlighted segment is visually similar to the selection.
This query mechanism facilitates detecting similar regions automatically, which is important for long sequences where a manual identification of similar time spans becomes more difficult. Identified time spans can be investigated, confirmed, and annotated with the respective label.

\section{Discussion}
Our examples showed that it is possible to achieve similar interpretations with the image-based 2D embedding of gaze spirals, as with clustering based on annotated data. 
This is an advantage for early analysis stages where no AOIs are available, because it helps identify groups of similar viewing behavior, which could be missed by statistical methods on the data as a whole.
Furthermore, the reasons \textit{why} scanpaths were grouped can be a investigated directly with the gaze spirals, reducing the frequency of time-consuming investigations in the recorded videos.
However, we identified some points that have to be considered for the application of gaze spirals in the future.

\paragraph{Scalability:}
Considering the temporal scalability of our approach, we tested gaze spirals on data with up to one hour of recording time. While this comprises a large number of possible experiments, it does not cover pervasive recordings over multiple hours. In such cases, we recommend reducing the sampling rate, as a trade-off between overview and representation of details. Individual fixations might become unrecognizable, but in the  context of long-term recording, it is typically more important to identify patterns of specific activities, which would still be possible, even with a reduced sampling rate. If a frame-by-frame representation is depicted, gaze spirals scale better than their linear counterpart.
Figure~\ref{fig:scalability} shows a comparison between a linear slitscan and the corresponding gaze spiral. It should be noted that the linear slitscan results in an image with a resolution of 367,122\,$\times$\,100 pixels, which is often not even a valid resolution for rendering libraries. We scaled the width of the slitscan down to 5\% of the original size before we could compare both visualizations. Given the same screen width, the gaze spiral is easier to interpret without extensive zooming. 

\paragraph{Interpretability:}
Visual patterns in gaze spirals mainly depict time spans when people where fixating a region (constant pattern) and changes of the point of regard (abrupt change in the pattern). The identification of similar patterns over time and between recordings is supported by queries. However, the interpretation and assignment of specific patterns to the respective AOI or activity requires analysts to investigate the underlying video, especially if a new pattern is analyzed. Hence, we implemented preview thumbnails that show on demand the corresponding video frame to help interpret patterns in the spiral.

\paragraph{Sequence Interpretation:}
We see the main purpose of gaze spirals in the compact representation of scanpath properties in the context of the stimulus. This means, we aim to overview the sequential order participants looked at AOIs, starting in the center of the spiral. With time progression, the spiral will grow and new timesteps appear on the outer arm of the spiral. Due to the inherent design of the gaze spiral, these new time spans will have more visual space than the inner parts. With this design, gaze spirals are also useful for live visualization, where current events are more important than older events. However, this design impairs a comparison of sequence durations. We could achieve this by decreasing $k$ depending on $t$. Due to general perceptual issues of people interpreting angular segments in comparison to linear segments, we would rather suggest including an additional view with selected linear segments, if the exact visual comparison of sequence durations is necessary. 

\paragraph{Head and Eye Movement}
A more general discussion relates to the question in which scenarios an egocentric video is sufficient and when eye tracking provides benefits. Our approach supports replacing the gaze position by a static scanline to account for scenarios with much head movement. 
We noticed that a slitscan from the egocentric video alone results in much smoother patterns while the gaze-based approach results in short patterns with numerous abrupt changes. However, in situations with multiple AOIs in the field of view, the gaze-supported slitscan provides much more details. A hybrid solution could provide smooth results while preserving important details, by combining the egocentric approach in situations with much movement (e.g., walking) with the gaze-based approach when people investigate a situation in detail.

\section{Conclusion and Future Work}

We presented a new approach to visualize scanpaths of multiple recordings from mobile eye tracking data. The resulting gaze spirals show pattern sequences that indicate in which order people looked at different AOIs and performed specific activities. Individual gaze spirals serve as glyphs for each recorded scanpath and are displayed in a 2D embedding to help identify and interpret groups of similar behavior based on image-features, without the necessity to annotate AOIs.
Note that similar to cluster analysis, our approach requires the user to be somewhat proficient with sequence alignment and 2D embeddings. One of our future goals is to make this process more accessible to the general, less proficient, user. At this time, we provide open access to our scripts for generating gaze spirals\footnote{https://github.com/Maurice189/gaze-spirals. Last checked on April 14, 2022.}.
With our technical evaluation, we showed that image-based calculations with established scanpath comparison methods lead to results comparable with annotated data. 

For future work, we further plan to conduct a user study on the annotation performance of gaze spirals in comparison other techniques such as the drawing and tracking of polygons. For the annotation of the art gallery dataset, we measured an average annotation time of $5$:$49$ minutes per recording, which is approximately twice the average run time of the videos.  
We see the proposed approach as a starting point for data exploration because it can be linked easily with detail views, e.g., gaze replays, while the overview is maintained. 
Thus, we think that gaze spirals are particularly relevant for hypothesis-generation purposes.
Due to the high scalability with respect to time, we further plan to investigate if the technique could handle long-term pervasive eye tracking data, which will become increasingly important in the years to come.
%which often comprises hours of recorded video data and 

\begin{acks}
This work was funded by the Deutsche Forschungsgemeinschaft (DFG, German Research Foundation) -- project ID 251654672 -- TRR 161 (project B01) and under Germany's Excellence Strategy – EXC 2120/1 – 390831618.
\end{acks}

\bibliographystyle{ACM-Reference-Format}
\bibliography{acmart}

\end{document}